Spin Polarized Conductance Induced by Tunneling Through
A Magnetic Impurity

D. Schmeltzer
Department of Physics, City College
of the City University of New York
New York, NY  10031
and
A. R. Bishop, A. Saxena, and D. L. Smith
Theoretical Division
Los Alamos National Laboratory
Los Alamos, NM  87545

Abstract

Using the zero mode method we compute the conductance of a wire consisting of a magnetic impurity coupled to two Luttinger liquid leads characterized by the Luttinger exponent $\alpha (\geq 1)$. We find for resonance conditions, in which the Fermi energy of the leads is close to a single particle energy of the impurity, the conductance as a function of temperature is $G \sim (e^2/h)(T/T_F)^{2(\alpha-2)}$, whereas for off-resonance conditions the conductance is $G \sim (e^2/h)(T/T_F)^{2(\alpha-1)}$. By applying a gate voltage and/or a magnetic field, one of the spin components can be in resonance while the other is off-resonance causing a strong asymmetry between the spin-up and spin-down conductances.

Quantum ballistic transport is crucial in understanding mesoscopic electronic devices. Examples include the single electron transistor[1] and the mystery of the point contact $(2e^2/h)$ 0.7 conductance[2]. In recent years it has become clear that the spin degree of freedom can play a critical role in quantum ballistic transport. The appealing possibility of realizing a spin-polarized transistor[3] has stimulated research in spin dependent transport[4,5]. In the quantum wire regime the presence of electron-electron (e-e) interactions gives rise to a spin-charge separated Luttinger liquid, where transport is controlled by the charge exponent[6]. In order to obtain spin dependent conductance in the quantum regime one must break the Luttinger liquid spin symmetry. There are several possible approaches to break the spin symmetry. In a recent publication, one of us[7] has shown that in the presence of e-e interactions and spin dependent Fermi velocities, $V_{F\uparrow} \neq V_{F\downarrow}$, transport is governed by spin dependent transport exponents which gives rise to a wire with spin polarized conductance.

In this Letter we propose a new mechanism leading to a spin polarized quantum wire. We consider a "wire" consisting of two leads coupled together through a magnetic impurity. Each lead is a one dimensional Luttinger liquid. The impurity has two electronic levels, one for each spin. The single particle levels in the impurity obey $|\varepsilon_\uparrow - \varepsilon_\downarrow| = \Delta$ (where $\Delta$ is the Zeeman energy). The left Luttinger liquid is confined to $-L \leq x \leq -a$ and the right one to $a \leq x \leq L$ where $L \gg a$. The magnetic impurity is confined to $|x| \leq a$. A schematic diagram of the structure is depicted in Fig. 1. Because of the Zeeman splitting the conductances $G_\uparrow$ and $G_\downarrow$ of the wire can be different. Such a wire can be realized in a semiconductor device structure in which the electron density of the leads can be varied by a gate voltage. The gate voltage can change the Fermi energy of the leads to achieve a situation where $|\varepsilon_\uparrow - E_F| \gg |\varepsilon_\downarrow - E_F| \to 0$. Transport across the wire is qualitatively different for the off-resonance $|\varepsilon_\uparrow - E_F| \gg \gamma$ and the resonance $|\varepsilon_\downarrow - E_F| \ll \gamma$ cases, where $\gamma$ is the matrix element coupling the leads to the impurity. (To be specific, we take spin-down to be in resonance and spin-up to be off-resonance.) We first present an intuitive discussion of the conductance in the two cases and then derive the results in detail.

For the off-resonance case we can replace the matrix elements $\gamma$ between the impurity and the leads by an effective "weak link" between the two leads with an effective coupling $t_\uparrow = \frac{\gamma^2}{|\varepsilon_\uparrow - E_F|}$. We thus obtain a weak link problem in the Luttinger liquid description with a coupling Hamiltonian: $t_\uparrow [C_\uparrow^+(a) C_\uparrow(-a) + \text{h.c.}]$. Following Luttinger theory[6], we find for the conductance, $G_\uparrow \sim \frac{e^2}{h} |t_\uparrow|^2 \left(\frac{T}{T_F}\right)^{2(\alpha-1)}$, where $\alpha$ is the Luttinger exponent $\alpha = \frac{1}{2}\left(\frac{1}{K_c} + \frac{1}{K_s}\right) \geq 1$; here $K_c$ and $K_s$ describe the charge- and spin-density-wave interactions, respectively. $\alpha = 1$ corresponds to the non-interacting case.

In the resonant case, the limit $|\varepsilon_\downarrow - E_F| \to 0$ allows us to replace the matrix elements $\gamma$ between the impurity and the leads by an effective coupling between the two leads, $t_\downarrow \sim \gamma^2$, that corresponds to tunneling between the two leads at different times. We find a new tunneling problem with long time correlations, $t_\downarrow \int_0^t dt_1 [C_\downarrow^+(-a,t) C_\downarrow(a,t_1) + h.c.]$. The additional time integration changes the scaling dimensions of $t_\downarrow$ by one. Due to the long time correlations, we find that the resonance condition gives rise to a shift in the tunneling exponent, $\alpha \to (\alpha - 1)$. The conductance for the resonant case becomes, $G_\downarrow \simeq \frac{e^2}{h} |t_\downarrow|^2 \left(\frac{T}{T_F}\right)^{2(\alpha-2)}$. As a result, the conductance for the off-resonance case is much smaller than that for the resonant case: $G_\uparrow \ll G_\downarrow$ leading to a spin polarized current, $I_\uparrow \ll I_\downarrow$. For temperature $T \to 0$ and $1 < \alpha < 2$, we obtain $G_\uparrow \to 0$ and $G_\downarrow \to \frac{e^2}{h}$.

We now quantify the above qualitative discussion and present our model and results in detail. The Hamiltonian for the wire can be split into three parts. $H = H^{leads} + H_i + H_T$, where $H^{leads} = H^L + H^R$ represents the left and right Luttinger liquids. The magnetic impurity is described by the Hamiltonian $H_i = \sum_{\sigma=\uparrow,\downarrow} \varepsilon_\sigma d_\sigma^+ d_\sigma$, where $\varepsilon_\uparrow \neq \varepsilon_\downarrow$, $\tilde{\varepsilon}_\sigma = \varepsilon_\sigma - \bar{\mu} < 0$, $\bar{\mu} = \frac{1}{2}(\mu_L + \mu_R)$, and $eV_{DS} = \mu_L - \mu_R$. Here, $\varepsilon_\sigma$ is the single particle energy in the impurity for spin $\sigma$ electrons, $\bar{\mu}$ is the chemical potential in the leads, and $V_{DS}$ is the voltage between the left and right leads. This Hamiltonian describes either an impurity in a magnetic field or a ferromagnetic impurity[8,9]. The tunneling Hamiltonian is given by $H_T = -\gamma \sum_{\sigma=\uparrow,\downarrow} [d_\sigma^+ C_\sigma(a) + d_\sigma^+ C_\sigma(-a)] + h.c.$

We integrate the impurity degrees of freedom and find, $d_\sigma(t) = \frac{i\gamma}{\hbar} \int_0^t dt_1 K_\sigma(t - t_1) [C_\sigma(-a, t_1) + C_\sigma(a, t_1)]$. Next, we substitute $d_\sigma(t)$ and $d_\sigma^+(t)$ into the tunneling Hamiltonian and obtain the time dependent tunneling Hamiltonian $\tilde{H}_T(t)$ between the two leads. The tunneling Hamiltonian $\tilde{H}_T(t)$ takes the form:

$$\tilde{H}_T(t) = -\frac{i\gamma^2}{\hbar} \sum_{\sigma=\uparrow,\downarrow} \int_0^t dt_1 \{K_\sigma(t-t_1)[C_\sigma^+(a,t)C_\sigma(-a,t_1) + C_\sigma^+(-a,t)C_\sigma(a,t_1) \\ + C_\sigma^+(a,t)C_\sigma(a,t_1) + C_\sigma^+(-a,t)C_\sigma(-a,t_1)] , \quad (1)$$

where $K_\sigma(t-t_1) = i[G_\sigma^>(t-t_1) - G_\sigma^<(t-t_1)]$. Here $G_\sigma^>(t-t_1)$ and $G_\sigma^<(t-t_1)$ are the advanced and retarded Green's function for the impurity. For an isolated impurity $K_\sigma(t-t_1) = e^{-i\omega_\sigma(t-t_1)}$ and $\omega_\sigma = \tilde{\varepsilon}_\sigma/\hbar$.

The final step in the calculation is to derive a continuum approximation. To do so we introduce a momentum cut-off $\Lambda$ for the fermions $C_\sigma(x)$, $C_\sigma^+(x)$ and integrate out states with momentum $|q| > \Lambda$. This integration induces a self energy for the single particle states in the impurity $\omega_\sigma \to \tilde{\omega}_\sigma + i\Gamma_\sigma$, where $\Gamma_\sigma = \Gamma_\sigma^L = \Gamma_\sigma^R$,
$\Gamma_\sigma^L = 2\gamma^2 \int_0^{K_F-\Lambda} dq\, \delta[\varepsilon_\sigma - E_\sigma^L(q)]$ and $\Gamma_\sigma^R = 2\gamma^2 \int_0^{K_F-\Lambda} dq\, \delta[\varepsilon_\sigma - E_\sigma^R(q)]$, where $E_\sigma^L(q), E_\sigma^R(q)$ are the single particle excitations in the leads far from the Fermi energy. (If $\tilde{\varepsilon}_\sigma \to 0$ and $q < K_F - \Lambda$ there is no solution, $\varepsilon_\sigma = E_\sigma^{R(L)}(q)$ giving rise to $\Gamma_\sigma = 0$).
We now consider $|\tilde{\varepsilon}_\downarrow| \simeq \hbar V_F \Lambda$, $\Lambda = \frac{1}{a}$ and $|\tilde{\varepsilon}_\downarrow| << \hbar V_F \Lambda$. This gives rise to two different scaling dimensions of the tunneling operator in Eq. 1. The fermion operators $\hat{C}_\sigma(x)$ restricted to momentum $|q| \leq \Lambda$ replace the bare fermion operators $C_\sigma(x)$.

At long wavelengths we replace the fermions $\tilde{C}_\sigma(x)$ by the bosonic representation. We use open boundary conditions for the right and left leads. Following Ref. 10, we have for the left lead $\hat{C}_\sigma(x = -L) = \hat{C}_\sigma(x = -a) = 0$. The bosonic representation for the fermions in the left lead is given by:

$$\hat{C}_\sigma(x<-a) = \frac{1}{\sqrt{2\pi a}} \chi_{L,\sigma} e^{i\alpha_\sigma}\left[ e^{i(\frac{\pi}{L}N_\sigma + K_F)x} e^{i\sqrt{4\pi}\theta_\sigma(x)} - e^{-i(\frac{\pi}{L}N_\sigma + K_F)x} e^{i\sqrt{4\pi}\theta_\sigma(-x)} \right]. \quad (2)$$

$\chi_{L,\sigma}$ is a real Majorana fermion, $\theta_\sigma(x)$ is the bosonic variable and $[\alpha_\sigma, N_{\sigma'}] = -i\delta_{\sigma,\sigma'}$ are the zero mode variables for the left lead. Following Ref. 10 we find the bosonic form of the Luttinger liquid in the left lead as: $H^L = H_L^{(n=0)} + H_L^{(n\neq 0)}$, $H_L^{(n=0)}$ is the zero mode part, $H_L^{(n=0)} = \frac{\hbar\pi}{4L}V_c(N_\uparrow + N_\downarrow)^2 + \frac{\hbar\pi}{L}V_s\left(\frac{N_\uparrow - N_\downarrow}{2}\right)^2$; and $H_L^{(n\neq 0)}$ is the bosonic part,
$H_L^{(n\neq 0)} = \int_{-L}^{L} dx \left[ V_c(\partial_x \tilde{\theta}_c)^2 + V_s(\partial_x \tilde{\theta}_s)^2 \right]$, where $\tilde{\theta}_c$ and $\tilde{\theta}_s$ represent the renormalized bosonic fields, $\theta_{c(s)}^{(x)} = \frac{K_{c(s)}^{-1/2}}{2}(\tilde{\theta}_{c(s)}^{(x)} - \tilde{\theta}_{c(s)}^{(-x)}) + \frac{K_{c(s)}^{-1/2}}{2}(\tilde{\theta}_{c(s)}^{(x)} + \tilde{\theta}_{c(s)}^{(-x)})$,
$\theta_c(x) = \frac{\theta_\uparrow(x) + \theta_\downarrow(x)}{\sqrt{2}}$, $\theta_s(x) = \frac{\theta_\uparrow(x) - \theta_\uparrow(x)}{\sqrt{2}}$ with $K_c < 1$, $K_s \approx 1$. $K_c$ and $K_s$ describe the interactions of the charge and spin excitation. $N_\uparrow$ and $N_\downarrow$ are the added electrons with respect to the Fermi energy. $N_\uparrow + N_\downarrow$ is the added charge and $\frac{N_\uparrow - N_\downarrow}{2}$ is the added spin. $0 \leq \alpha_\sigma \leq 2\pi$ is the zero mode phase conjugated to $N_\sigma$. $\theta_c^{(x)}$ and $\theta_s^{(x)}$ are bare charge and spin particle hole excitations. $H_L^{(n \neq 0)}$ represent the bosonic charge and spin density wave excitations with charge and spin velocities $V_c$ and $V_s$. For the right lead $a \leq x \leq L$ we also use open boundary conditions, $\hat{C}_\sigma(x = L) = \hat{C}_\sigma(x = a) = 0$:

$$\hat{C}_\sigma(x > a) = \frac{1}{\sqrt{2\pi a}} \chi_{R,\sigma} e^{i\beta_\sigma} \left[ e^{i(\frac{\pi}{L}n_\sigma + K_F)x} e^{i\sqrt{4\pi}\psi_\sigma(x)} - e^{-i(\frac{\pi}{L}n_\sigma + K_F)x} e^{i\sqrt{4\pi}\psi_\sigma(-x)} \right], \quad (3)$$

where $\psi_\sigma(x)$ is the bosonic field in the right lead (the equivalent of $\theta_\sigma(x)$), and $\beta_\sigma$ $n_\sigma$ are the zero modes variables, $[\hat{\beta}_\sigma, n_{\sigma'}] = -i\delta_{\sigma,\sigma'}$. $\beta_\sigma$ is the equivalent of $\alpha_\sigma$ and $n_\sigma$ is the equivalent of $N_\sigma$. For simplicity, we assume that the right lead is identical to the left one. The Hamiltonian for the right lead obeys: $H_R^{(n=0)}(n_\uparrow; n_\downarrow) \equiv H_L^{(n=0)}(N_\uparrow = n_\uparrow; N_\downarrow = n_\downarrow)$ and $H_R^{(n \neq 0)}(\psi_\uparrow; \psi_\downarrow) = H_L^{(n \neq 0)}(\theta_\uparrow = \psi_\uparrow; \theta_\downarrow = \psi_\downarrow)$.

We substitute the bosonic representation given by Eqs. 2 and 3 into Eq. 1 and find the effective tunneling Hamiltonian $h_T(t)$; between the leads:

$$h_T(t) = -i\lambda \sum_{\sigma=\uparrow,\downarrow} \chi_{R,\sigma}\chi_{L,\sigma} \int_0^t dt_1 \{K_\sigma(t - t_1)[e^{-i\hat{\psi}_\sigma(t)}e^{i\hat{\theta}_\sigma(t_1)} - e^{-i\hat{\theta}_\sigma(t)}e^{i\hat{\psi}_\sigma(t_1)}] - h.c.\}. \quad (4)$$

In Eq. (4) we have used the notations; $\hat{\theta}_\sigma = \alpha_\sigma + \sqrt{4\pi}\theta_\sigma$, $\hat{\psi}_\sigma \equiv \beta_\sigma + \sqrt{4\pi}\psi_\sigma$, $\lambda \equiv (2\gamma)^2 \frac{2\sin(K_F^R a) \sin(K_F^L a)}{ah}$, and $K_F^R = K_F^L$.

The tunneling current computed within the zero mode formulation is given by $I_\sigma = e\frac{d\hat{n}_\sigma}{dt} = -e\frac{d\hat{N}_\sigma}{dt} = \frac{1}{2}e\frac{d\hat{J}_\sigma}{dt}$, and $\hat{J}_\sigma = \hat{n}_\sigma - \hat{N}_\sigma$. This $\hat{J}_\sigma = \hat{n}_\sigma - \hat{N}_\sigma$ is expressed in terms of $J_\sigma = n_\sigma - N_\sigma \equiv i\left(\frac{d}{d\alpha_\sigma} - \frac{d}{d\beta_\sigma}\right) \equiv 2i\frac{d}{d\gamma_\sigma}$, where $\gamma_\sigma \equiv \alpha_\sigma - \beta_\sigma$, $[\gamma_\sigma, J_\sigma] = -2i$. $J_\sigma$ is the current operator in the Schrödinger picture and $\hat{J}_\sigma$ is the current operator in the Heisenberg picture. Using the interaction picture we express the current operator $\hat{J}_\sigma(t)$ in terms of the Schrödinger current, $J_\sigma$

$$<<\hat{J}_\sigma(t)>> = << T_{c_t}\left\{\exp\left[\frac{-i}{\hbar}\int_{c_t} h_T(t_1)dt_1\right]\right\}J_\sigma >> \quad (5)$$

In Eq. (5) $<<\ >>$ stands for the thermodynamic average at temperature T with respect to the zero-mode Hamiltonian: $H_L^{(n=0)} + H_R^{(n=0)} + \mu_L(N_\uparrow + N_\downarrow) + \mu_R(n_\uparrow + n_\downarrow)$, $eV_{DS} \equiv \mu_L - \mu_R$ and bosonic part $H_R^{(n \neq 0)} + H_L^{(n \neq 0)}$. Here $T_{c_t}$ indicates the time order and $C_t$ is the contour in the Keldysh representation[11,12]. We find to second order in $\lambda^2 \propto \gamma^4$ from Eq. (5) that the current $I_\sigma = \frac{e}{2} << \frac{d\hat{J}_\sigma(t)}{dt} >>$ is given by

$$I_\sigma = e\left(\frac{-i}{\hbar}\right)^2 i << \int_0^t dt_1 \left\{ h_T(t_1 - i\varepsilon) \frac{d}{d\gamma_\sigma} h_T(t) - \frac{1}{2}\left(\frac{d}{d\gamma_\sigma} h_T(t)\right)\left[h_T(t_1 - i\varepsilon) + h_T(t_1 + i\varepsilon)\right]\right\} >> \tag{6}$$

Equation 6 is our result for the tunneling current expressed in terms of the zero mode derivative of the tunneling Hamiltonian. The excitation values in Eq. 6 depend on the bosonic correlation function:

$$<< e^{i\sqrt{4\pi}\theta_{c(s)}^{(t)}} e^{-i\sqrt{4\pi}\theta_{c(s)}^{(t_1)}} >> = \left\{ \frac{i\frac{\pi V_F}{L_T \Lambda}}{\sinh\left[\frac{\pi V_F}{L_T \Lambda}(t - t_1)\right]} \right\}^{\frac{1}{K_{c(s)}}},$$

where $L >> L_T = \frac{\hbar V_F}{K_B T}$ is the thermal length; similar correlations exist for the $\psi_\sigma$ fields. From the zero mode part (see Ref. 10) we find in the limit $L >> L_T$,

$$<< n_\uparrow + n_\downarrow - N_\uparrow - N_\downarrow >> = \frac{eV_{DS}}{h}\left(\frac{2L}{V_c}\right) \text{ and } << e^{i\alpha_\sigma(t)} e^{-i\alpha_\sigma(t_1)} >>$$

$$= \exp\left(i\frac{\omega_{DS}}{\hbar}(t - t_1)\right), \quad \omega_{DS} \equiv 2\pi \frac{eV_{DS}}{\hbar}.$$

We first evaluate the non-resonant case, $|\tilde{\varepsilon}_\uparrow| = |\varepsilon_\uparrow - E_F| \geq \gamma$. This allows us to compute the spin up current $I_\uparrow$. Using Eq. (6) we find

$$I_\uparrow = \frac{e}{\hbar^2}\left[\frac{\tilde{\lambda}^2}{((\varepsilon_\uparrow - \bar{\mu})/\hbar)^2 + \Gamma_\uparrow^2}\right] \int_0^t dt_1 2i\sin\omega_{DS}(t - t_1)\left[R(t - t_1 + i\varepsilon) - R(t - t_1 - i\varepsilon)\right], \tag{7}$$

where $R(t - t_1 \pm i\varepsilon) = \left[\frac{\frac{\pi V_F}{L_T \Lambda}}{\sinh\left[\frac{\pi V_F}{L_T \Lambda}(t - t_1 \pm i\varepsilon)\right]}\right]^{2\alpha}$ with $2\alpha \equiv 2\left(\frac{1}{2K_c} + \frac{1}{2K_s}\right)$. Equation

(7) is obtained after the impurity degrees of freedom have been integrated out. We obtain a weak link problem in a Luttinger liquid framework. We perform the integral in Eq. (7) for the linear voltage regime, $V_{DS} \to 0$, and find for the conductance, $G_\uparrow = I_\uparrow / V_{DS}$:

$$G_\uparrow = \frac{e^2}{\hbar}\left[\frac{4\tilde{\lambda}^2}{((\varepsilon_\uparrow - \bar{\mu})/\hbar)^2 + \Gamma_\uparrow^2}\right] \frac{\pi\cos(\pi\alpha)}{\Gamma(2\alpha)}\left(\frac{T}{T_F}\right)^{2(\alpha - 1)}, \tag{8}$$

where $b_T = L_T \Lambda \equiv \frac{T_F}{T} >> 1$. (The coupling constant $\lambda$ is related to the dimensionless coupling constant $\hat{\lambda}$, $\lambda \equiv \hat{\lambda}\hbar(V_F \Lambda)^2 = \tilde{\lambda}\hbar(V_F \Lambda)$ and $\Gamma(2\alpha)$ is the gamma function.)

The result in Eq. (8) shows that for $\alpha > 1$ the conductance decreases at low temperature. Except from the resonance factor, Eq. (8) is the Luttinger liquid result obtained for a weak link[6]. For this case we can use the weak link scaling theory, $\frac{d\hat{\lambda}}{ds} = (1-\alpha)\hat{\lambda}$, $e^s > b_T = \frac{\hbar V_F \Lambda}{K_B T}$. For $T \to 0$ and $\alpha > 1$ we obtain $G_\uparrow \to 0$. This means that the spin up current $I_\uparrow$ is completely backscattered by the impurity.

Next we consider the resonance case $\tilde{\varepsilon}_\downarrow \to 0$. In this limit $K_\downarrow(t-t_1) \sim 1$. As a result in evaluating Eq. (6) we find that the singular integrals are shifted from $(t-t_1)^{-2\alpha}$ to $(t-t_1)^{-2(\alpha-1)}$ in agreement with "long time" correlation. This gives the conductance $G_\downarrow = \frac{I_\downarrow}{V_{DS}}$ as

$$G_\downarrow = \frac{e^2}{\hbar} 4\hat{\lambda}^2 \frac{\pi \cos \pi \alpha}{(1-\alpha)(2-\alpha) \Gamma[2(\alpha-1)]} \left(\frac{T}{T_F}\right)^{2(\alpha-2)}, \qquad (9)$$

where $\alpha > 1$ and $\Gamma[2(\alpha-1)]$ is the gamma function. The results in Eq. (9) show that the current $I_\downarrow$ increases at low temperatures if $1 < \alpha < 2$.

The spin polarization of the current depends on the ratio $G_\uparrow/G_\downarrow \sim \left(\frac{T}{T_F}\right)^2$, which decreases at low temperatures giving perfect spin polarization $P = \frac{G_\downarrow - G_\uparrow}{G_\downarrow + G_\uparrow} \to 1$. Therefore the magnetic impurity acts as a spin polarizer. The result in Eq. (9) is in agreement with the scaling equation, $\frac{d\lambda}{ds} = (2-\alpha)\lambda$. The change in the scaling dimension of $\hat{\lambda}$, (as compared to the Luttinger case) is due to the long time correlation which induced an additional time integration. Thus for $1 < \alpha < 2$, $\lambda$ increases at long distances, so that at $T \to 0$ we have perfect transmission for $I_\downarrow$ with a conductance $G_\downarrow \to \frac{e^2}{h}$.

In order to understand why $G_\downarrow \to \frac{e^2}{h}$, as $T \to 0$, we present a scaling argument. We consider that the hopping constant in the leads is $t_0 \sim V_F$ and the tunneling matrix element between the impurity and leads is $\gamma \ll t_0$. The value of the conductance $G_\downarrow$ is determined by two scaling regions:

I. $1 \leq b \leq b_0 = \frac{\Lambda V_F}{\tilde{\omega}_\downarrow}$, $b \equiv e^s > 1$, $\tilde{\omega}_\downarrow \equiv |\varepsilon_\downarrow - E_F|/\hbar$. In this region $\tilde{\omega}_\downarrow(b) = \tilde{\omega}_\downarrow b$, $\lambda(b) = \lambda b^{(2-\alpha)}$, where $\lambda \sim \gamma^2$, $t_\downarrow(b) = \frac{\lambda(b)}{\tilde{\omega}_\uparrow(b)}$. For $b \leq b_0$, $t_\downarrow(b)$ increases.

II. $b > b_0$, $t_\downarrow(b) \equiv \frac{\lambda(b)}{\tilde{\omega}_\downarrow(b)}$, $t_\downarrow(b) = t_\downarrow(b_0)\left(\frac{b}{b_0}\right)^{1-\alpha} \xrightarrow[b\to\infty]{} 0$.

If $b = b_0 = \frac{\Lambda V_F}{\tilde{\omega}_\downarrow}$ such that $t_\downarrow(b_0) = t_0$ ($t_0$ is the matrix element in the leads), we have perfect transmission, $G_\downarrow \sim \frac{e^2}{h}$. The condition $t_\downarrow(b_0) = t_0$ determines the range of $\tilde{\omega}_\downarrow \equiv |\varepsilon_\downarrow - E_F|/\hbar$ for which one has perfect transmission. We find that perfect transmission is achieved for frequencies $\tilde{\omega}_\downarrow$ which obey $\tilde{\omega}_\downarrow \leq \left\{\left(\frac{\gamma}{t_0}\right)\gamma(V_F\Lambda)^{(1-\alpha)}\right\}^{\frac{1}{2-\alpha}}$.

Finally, we present a simple numerical estimate of the important parameters. A crucial parameter is the coherence length $L_\phi(T)$ which must be larger than the length of the wire L and the thermal length $L_T$. Using diffusion theory we have $L_\phi(T) = \sqrt{L_T \ell} > L \geq L_T$, where $\ell$ is the mean free path. In GaAs we have $V_F \approx 10^5 \text{m/sec}$, $\ell \approx 10^{-5}\text{m}$, $\lambda_F \approx 10^{-7}\text{m}$ (the Fermi wavelength) and the interaction parameter $\alpha \approx 1.2$. We find that the condition $L_\phi(T) > L \geq L_T$ is satisfied for $L \approx L_T \approx 10^{-6}\text{m}$. This estimate is for temperatures $T \approx 2$ K. Using these parameters we find that in the resonance case $\left(\frac{T}{T_F}\right)^{-1.6} = \left(\frac{\lambda_F}{L_T}\right)^{-1.6} \approx 10^{1.6}$, and for the off-resonance case $\left(\frac{T}{T_F}\right)^{0.4} = \left(\frac{\lambda_F}{L_T}\right)^{0.4} \approx 10^{-0.4}$. Thus the ratio of resonance to off-resonance conductance is large, about 100.

In conclusion, we have shown that a wire consisting of two Luttinger liquid leads coupled together through a magnetic impurity can act as a spin polarizing structure. The magnetic impurity breaks the spin symmetry of the Luttinger liquid. This effect requires magnetic impurities with different energy levels for the two spin orientations. A physical realization could be a semiconductor device structure with gate voltage dependent electron density in the leads coupled to a ferromagnetic quantum dot. The gate voltage can change the Fermi energy of the leads so that one spin orientation is resonant and the other off-resonant. Transport across the wire is then qualitatively different for the resonance and off-resonance cases. For the resonance case a long-time correlation is induced by the impurity. This correlation is reflected by a shift in the Luttinger exponent $\alpha \to (\alpha - 1)$. Consequently, the two conductances $G_\uparrow$ and $G_\downarrow$ are different. At low temperatures the off-resonant conductance goes to zero and the resonant conductance approaches the quantum conductance limit. In this limit the wire is a perfect spin filter.

This work was supported by the U. S. Department of Energy Grant No. DE-FG02-01ER43909. Work at Los Alamos was supported by the Los Alamos LDRD program.

Figure 1

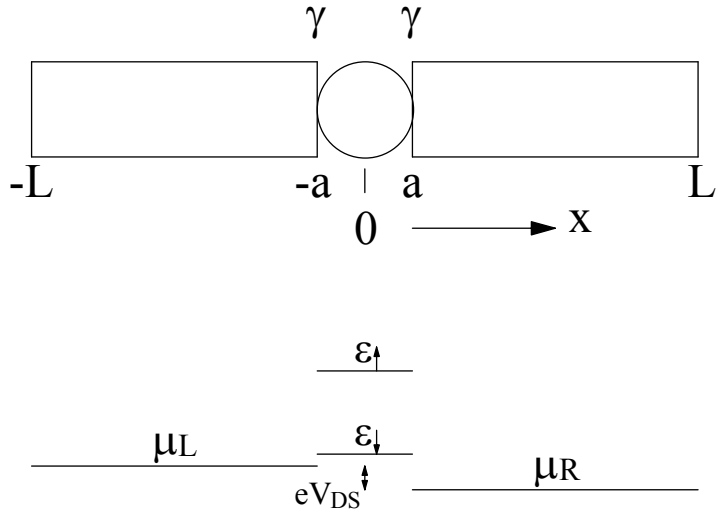

Figure Caption

Fig. (1) Schematic illustration of: (top panel) the structure consisting of two Luttinger liquid leads, extending between (-L and -a) and (a and L), connected by a hopping matrix element $\gamma$ to a magnetic impurity extending between (-a and a); and (bottom panel) energy level diagram showing the chemical potential of the left ($\mu_L$) and right ($\mu_R$) leads and the energy levels of the magnetic impurity. $V_{DS}$ is the voltage applied between the two leads.